\documentclass{jfm}
\usepackage{graphicx}
\usepackage{amsmath}
\usepackage{overpic}
\usepackage{float}
\usepackage{epstopdf,epsfig}
\usepackage{color,contour}
\usepackage{amssymb}
\usepackage[dvipsnames]{xcolor}
\usepackage{graphicx}
\usepackage{tabularx}
\usepackage{tabularx}
\usepackage[dvipsnames]{xcolor}
\usepackage[colorlinks,citecolor = blue, linkcolor=red,hyperindex,CJKbookmarks]{hyperref}
\usepackage[dvipsnames]{xcolor}
\usepackage{booktabs}

\usepackage{longtable}
\usepackage{enumerate}

\shorttitle{Regime transitions in high-$Ra$ vertical convection}
\shortauthor{Q. Wang et al.}

\title{Regime transitions in thermally driven high-Rayleigh number vertical convection }

\author{Qi Wang\aff{1,2},
Hao-Ran Liu\aff{1},
Roberto Verzicco\aff{1,3,4},
Olga Shishkina\aff{5}\corresp{\email{Olga.Shishkina@ds.mpg.de}},
and
Detlef Lohse\aff{1,5}\corresp{\email{d.lohse@utwente.nl}}
}
\affiliation{\aff{1}Physics of Fluids Group and Max Planck Center for Complex Fluid Dynamics, MESA+ Institute and J. M. Burgers Centre for Fluid Dynamics, University of Twente, P.O. Box 217, 7500AE Enschede, The Netherlands
\aff{2}Department of Modern Mechanics, University of Science and Technology of China, 230027 Hefei, China
\aff{3}Dipartimento di Ingegneria Industriale, University of Rome ``Tor Vergata”, Via del Politecnico 1, Roma 00133, Italy
\aff{4}Gran Sasso Science Institute - Viale F. Crispi, 767100 L'Aquila, Italy
\aff{5}Max Planck Institute for Dynamics and Self-Organization, 37077 G\"ottingen, Germany
}

\begin{document}

\maketitle

\begin{abstract}

\noindent Thermally driven vertical convection (VC) -- the flow in a box heated on one side and cooled on the other side, is investigated using direct numerical simulations over a wide range of Rayleigh numbers $10^7\le Ra\le10^{14}$ with fixed Prandtl number $Pr=10$, in a two-dimensional convection cell with unit aspect ratio. It is found that the dependence of the mean vertical centre temperature gradient $S$ on $Ra$ shows three different regimes: In regime I ($Ra \lesssim 5\times10^{10}$), $S$ is almost independent of $Ra$; In the newly identified regime II ($5\times10^{10} \lesssim Ra \lesssim 10^{13}$), $S$ first increases with increasing $Ra$ (regime ${\rm{II}}_a$), reaches its maximum and then decreases again (regime ${\rm{II}}_b$); In regime III ($Ra\gtrsim10^{13}$), $S$ again becomes only weakly dependent on $Ra$, being slightly smaller than in regime I. The transition from regime I to regime II is related to the onset of unsteady flows due to the ejection of plumes from the sidewall boundary layers. The maximum of $S$ occurs when these plumes are ejected over about half of the area (downstream) of the sidewalls. The onset of regime III is signatured with the appearance of layered structures near the top and bottom horizontal walls. The flow in regime III is characterized by a well-mixed bulk region due to continuous ejection of plumes over large fractions of the sidewalls, and due to the efficient mixing, the mean temperature gradient in the centre $S$ is smaller than that of regime I. In the three different regimes, significantly different flow organizations are identified: In regime I and regime ${\rm{II}}_a$, the location of the maximal horizontal velocity is close to the top and bottom walls; However, in regime ${\rm{II}}_b$ and regime III, banded zonal flow structures develop and the maximal horizontal velocity now is in the bulk region.

The different flow organizations in the three regimes are also reflected in the scaling exponents in the effective power law scalings $Nu\sim Ra^\beta$  and $Re\sim Ra^\gamma$. In regime I, the fitted scaling exponents ($\beta\approx0.26$ and $\gamma\approx0.51$) are in excellent agreement with the theoretical predication of $\beta=1/4$ and $\gamma=1/2$ for laminar VC (Shishkina, {\it{Phys. Rev. E.}} 2016, 93, 051102). However, in regimes II and III, $\beta$ increases to a value close to 1/3 and $\gamma$ decreases to a value close to 4/9.  The stronger $Ra$ dependence of $Nu$ is related to the ejection of plumes and larger local heat flux at the walls. The mean kinetic dissipation rate also shows different scaling relations with $Ra$ in the different regimes.

\end{abstract}

\begin{keywords}

\end{keywords}

\section{Introduction}

Thermally driven convective fluid motions are ubiquitous in various geophysical and astrophysical flows and are important in many industrial applications.  Rayleigh--B\'enard convection (RBC)\citep{ahlers2009heat,lohse2010small,chilla2012new,xia2013current}, where a fluid layer in a box is heated from below and cooled
from above, and vertical convection (VC) \citep{ng2015vertical,shishkina2016momentum,ng2017changes,ngbulk}, where  the fluid is confined between two differently heated isothermal vertical
walls, have served as two classical model problems to study thermal convection. VC was also called convection in a differentially heated vertical box in many early papers \citep{paolucci1989transition,le1998onset}. Both RBC and VC can be viewed as extreme cases of the more general so called tilted convection \citep{guo2015effect,shishkina2016thermal,wang2018flow,wang2018multiple,zwirner2018confined,zwirner2020influence}, with the tilt angle of $0^\circ$ for RBC and $90^\circ$ for VC.  We focus on VC in this study. VC finds many applications in engineering: thermal insulation using double-pane windows or double walls, horizontal heat transport in water pools with heated/cooled sideewalls , crystal growth procedures, nuclear reactors, ventilation of rooms, and cooling of electronic devices, to name only a few. VC has also served as a model to study thermally driven atmospherical circulation \citep{hadley1735vi,lappa2009thermal} or thermally driven circulation in the ocean, e.g., next to a ice-block \citep{thorpe1969effect,tanny1988dynamics}

The main control parameters in VC are the Rayleigh number $Ra\equiv g\alpha L^3\Delta/(\nu \kappa)$ and the Prandtl number $Pr\equiv \nu/\kappa$. Here, $\alpha$, $\nu$, $\kappa$ are the thermal expansion coefficient, the kinematic viscosity, and the thermal diffusivity of the convecting fluid, respectively, $g$ the gravitational acceleration, $\Delta \equiv T_h-T_c$ the temperature difference between the two side walls, and $L$ the width of the convection cell. The aspect ratio $\Gamma \equiv H/L$ is defined as the ratio of height $H$ over width $L$ of the domain. The responses of the system are characterized by the Nusselt number $Nu\equiv{QL}/{(k\Delta})$ and the Reynolds number $Re\equiv{UL}/{\nu}$, which indicate the non-dimensional heat transport and flow strength in the system, respectively. Here $Q$ is the heat flux crossing the system and $U$ the characteristic velocity of the flow.

Since the pioneering work of Batchelor \citep{batchelor1954heat}, who first addressed the case of the steady state heat transfer across double-glazed windows, VC has drawn significant attention especially in the 1980s and 1990s, and most of these studies used experiments or two-dimensional (2-D) direct numerical simulations (DNS) in a square domain with unit aspect ratio. For relatively low $Ra$ (e.g. $Ra < 10^3$), the flow is weak and heat is transferred mainly by thermal conduction. With increasing $Ra$, typical stratified flow structures appear in the bulk region \citep{de1983natural}, while the flow is still steady. With a further increase in $Ra$, the flow becomes unsteady with periodical/quasi-periodical  or chaotic motions \citep{paolucci1989transition,le1998onset}, and eventually turbulent when $Ra$ is high enough \citep{paolucci1990direct}.

The onset of unsteadiness has been well explored in the past  \citep{chenoweth1986natural,paolucci1989transition,janssen1995influence,le1998onset}. \citet{paolucci1989transition} investigated the influence of the aspect ratio $\Gamma$ on the onset of unsteadiness for 2-D VC with $Pr=0.71$. They found that for $\Gamma \gtrsim 3$ the first transition from the steady state is due to an instability of the sidewall boundary layers, while for smaller aspect ratios $0.5\le\Gamma \lesssim 3$, it is due to internal waves near the departing corners. Such oscillatory instability due to internal waves was first pointed out by \citet{chenoweth1986natural}. \citet{paolucci1989transition}  also found that for $\Gamma=1$, the critical Rayleigh number $Ra_c$ for the onset of unsteadiness lies between $1.8\times10^{8}$ and $2\times10^8$. Later work with $Pr=0.71$ and $\Gamma=1$ by \citet{le1998onset} also showed that the internal gravity waves play an important role in the time-dependent dynamics of the solutions, and $1.81\times10^8\le Ra_c\le1.83\times10^8$ was found for the range of the critical Rayleigh number. \citet{janssen1995influence} studied the influence of $Pr$ on the instability mechanisms for $\Gamma=1$, and found that for $0.25\le Pr\le2$, the transition occurs through periodic and quasi-periodic flow regimes. One bifurcations is related to an instability occurring in a jet-like fluid layer exiting from the corners of the cavity where the vertical boundary layers are turned horizontal. Such jet-like flow structures are responsible for the generation of internal gravity waves \citep{chenoweth1986natural,paolucci1989transition}. The other bifurcation occurs in the boundary layers along the vertical walls. Both of these instabilities are mainly shear-driven. For $2.5\le Pr\le7$, \citet{janssen1995influence} found an ``immediate'' (i.e., sharp) transition from the steady to the chaotic flow regime, without intermediate regimes. This transition is also caused by boundary layer instabilities. They also showed that $Ra_c$ significantly increases with increasing $Pr$, e.g. for $Pr=4$, the flow can still be steady with $Ra=2.5\times10^{10}$. However, due to the computation limit, unsteady motions for the large-$Pr$ cases were largely unexplored in the past.

Also the flow structures for VC were examined in detail. A typical flow feature for VC is the stably-stratified bulk region \citep{de1983natural,ravi1994high,trias2007direct,chong2020caf}. Such stratification can be quantified by the time-averaged non-dimensional temperature gradient at the centre, namely
\begin{gather}
S\equiv \left<(L/\Delta)(\partial{T}/\partial{z})_c\right>_t. \label{temg}
\end{gather}

\noindent Here $\left<\right>_t$ denotes time average. \citet{gill1966boundary} derived asymptotic solutions for high $Pr$, and predicted $S=0.42$ as $Ra\to\infty$, while an accurate solution of the same system by \citet{blythe1983thermal} predicts the value 0.52. Later DNS results for $Ra=10^8$, $Pr=70$ yield $S=0.52$ \citep{ravi1994high}, which is in excellent agreement with the theoretical predication by \citet{blythe1983thermal}. However, for small $Pr$, the structure of the core and the vertical boundary layer are no longer similar to those predicted by the asymptotic solutions which are valid for large $Pr$ \citep{blythe1983thermal}. And unfortunately, there exists no such asymptotic theory for finite $Pr$. Only \citet{graebel1981influence} has presented some approximate solutions, in which he has neglected some terms in the equations. For $Pr=0.71$, his predication yields $S=0.49$, which is considerably smaller than the value $S\approx1$ from DNS \citep{ravi1994high,trias2007direct}. It was concluded that $S$ is independent of $Ra$ for $Ra\le10^{10}$ \citep{paolucci1990direct}, however, it is evident that the dependence of $S$ on $Ra$ and $Pr$, especially for those with high $Ra>10^{10}$ and low $Pr<0.71$, are still poorly understood.

A key question in the study of thermal convection is: How do $Nu$ and $Re$ depend on $Ra$ and $Pr$? This question has been extensively addressed in RBC over the past years  \citep{ahlers2009heat}. For RBC, the mean kinetic dissipation rate ($\epsilon_u$) and  thermal dissipation rate ($\epsilon_\theta$) obey exact global balances, featuring $Ra$, $Nu$ and $Pr$ \citep{shraiman1990heat}. In this problem, in a series of papers, \cite{grossmann2000scaling,grossmann2001thermal,grossmann2002prandtl,grossmann2004fluctuations} developed a unifying theory to account for $Nu(Ra,Pr)$ and $Re(Ra,Pr)$ over wide parameter ranges. The central idea of the theory is a decomposition of $\epsilon_u$ and $\epsilon_\theta$ into their boundary layer and bulk contributions. The theory has been well confirmed through various  experiments and numerical simulations \citep{stevens2013unifying}. This theory has also been applied to horizontal convection \citep{shishkina2016heat,shishkina2016prandtl} and internally heated convection \citep{wang2020ihc}. However, in VC, the exact relation for $\epsilon_u$ does not hold, which impedes the applicability of unifying theory to the scalings in VC \citep{ng2015vertical}.

As compared to RBC, in VC much less work was devoted to the dependences $Nu(Ra,Pr)$ and $Re(Ra,Pr)$. Past studies suggested  power law dependences, i.e., $Nu\sim Ra^\beta$ and $Re\sim Ra^\gamma$, at least in a certain $Ra$-range. The reported scaling exponent $\beta$  was found to vary from 1/4 to 1/3 \citep{le1998onset,xin1995direct, trias2007direct, trias2010direct,ng2015vertical,shishkina2016momentum,wang2019non,ng2020non}, depending on the  $Ra$-range and $Pr$. \cite{ng2015vertical} simulated three-dimensional (3-D) VC with periodic conditions in the range $10^5\le Ra\le10^9$ with $Pr=0.709$, and obtained $\beta=0.31$ for the considered range. For much larger $Pr\gg1$ and using laminar boundary layer theories, \cite{shishkina2016momentum} theoretically derived $Nu\sim Ra^{1/4}$ and $Re\sim Ra^{1/2}$  These theoretical results are in excellent agreement with direct numerical simulations for $Ra$ from $10^5$ to $10^{10}$ in a cylindrical container with aspect ratio $\Gamma=1$. The power law exponents $\beta=1/4$ and $\gamma=1/2$ were also confirmed by the DNS of \citet{ng2020non} in a 3-D cell with span-wise periodic boundary conditions for $10^8\le Ra\le1.3\times10^9$. For 2-D VC, past studies with $Ra\le10^{10}$ also showed that $\beta$ is closer to 1/4 than 1/3 \citep{xin1995direct,trias2007direct,trias2010direct,wang2019non}. \cite{wang2019non} simulated 2-D VC over $10^5\le Ra\le10^9$ for fixed $Pr=0.71$, they found $\beta\approx0.27$ and $\gamma\approx0.50$.

Most of the simulations for VC were conducted for $Ra\lesssim10^{10}$. The high-$Ra$ simulations become stiff owing to a decrease in boundary-layer thicknesses with increasing $Ra$. As a result, little is known about what will happen at $Ra$ much larger than $10^{10}$. In this study, we try to fill this gap by performing DNS up to $Ra=10^{14}$. The price we have to pay is that for those large $Ra$ we have to restrict us to 2-D. However, in \cite{van2013comparison} 2-D and 3-D simulations for RBC were compared in detail, and many similarities were found for $Pr\ge1$. The unifying theory is also 2-D, but it works well in 3-D \citep{stevens2013unifying}. So we think 2-D and 3-D results share similar features.

The main questions we want to address in this study are:
\begin{itemize}
\item [(i)]Is the conclusion that $S$ is independent of $Ra$ for $Ra\le10^{10}$ \citep{paolucci1990direct} still valid for $Ra$ much larger than $10^{10}$?

\item [(ii)] How does the global flow organization (mean temperature and velocity profiles) change with increasing $Ra$ up to $10^{14}$?

\item [(iii)] How robust are the laminar scaling relations $Nu\sim Ra^{1/4}$ and $Re\sim Ra^{1/2}$ \citep{shishkina2016momentum} for higher $Ra$? Will new scaling relations appear for $Ra$ much larger than $10^{10}$?

\end{itemize}

 We find that $S$ is not independent of $Ra$ over the studied parameter range at all. Instead, we find that apart from the small-$Ra$ regime (now called regime I), where $S$ only weakly depends on $Ra$ \citep{paolucci1990direct}, there are further regimes for $Ra\gtrsim5\times10^{10}$ with different scaling relations. In regime II ($5\times10^{10} \lesssim Ra \lesssim 10^{13}$), with increasing $Ra$, $S$ first increases (regime ${\rm{II}}_a$) to its maximum and then decreases (regime ${\rm{II}}_b$) again. In regime III ($Ra\gtrsim10^{13}$), $S$ again becomes weakly dependent on $Ra$, with a smaller value than that of regime I.  Furthermore, we find that the laminar power law exponents $\beta=1/4$, $\gamma=1/2$ undergo sharp transitions to $\beta\approx1/3$ and $\gamma\approx4/9$ when $Ra\gtrsim5\times10^{10}$, i.e. at the transition from regime I to regime II.

 The rest of the paper is organised as follows. Section \ref{sec2} describes the governing equations and numerical methods. The different  flow organizations in the different regimes are studied in \S \ref{sec3}. Finally in \S \ref{sec4}, we discuss the transition of the scaling relations for of heat and momentum transport between the different regimes. Finally \S \ref{sec5} contains a summary and an outlook.

\section{Numerical procedures}\label{sec2}

\begin{figure}
  \centering
  \vskip 2mm
  \begin{overpic}[width=0.5\textwidth]{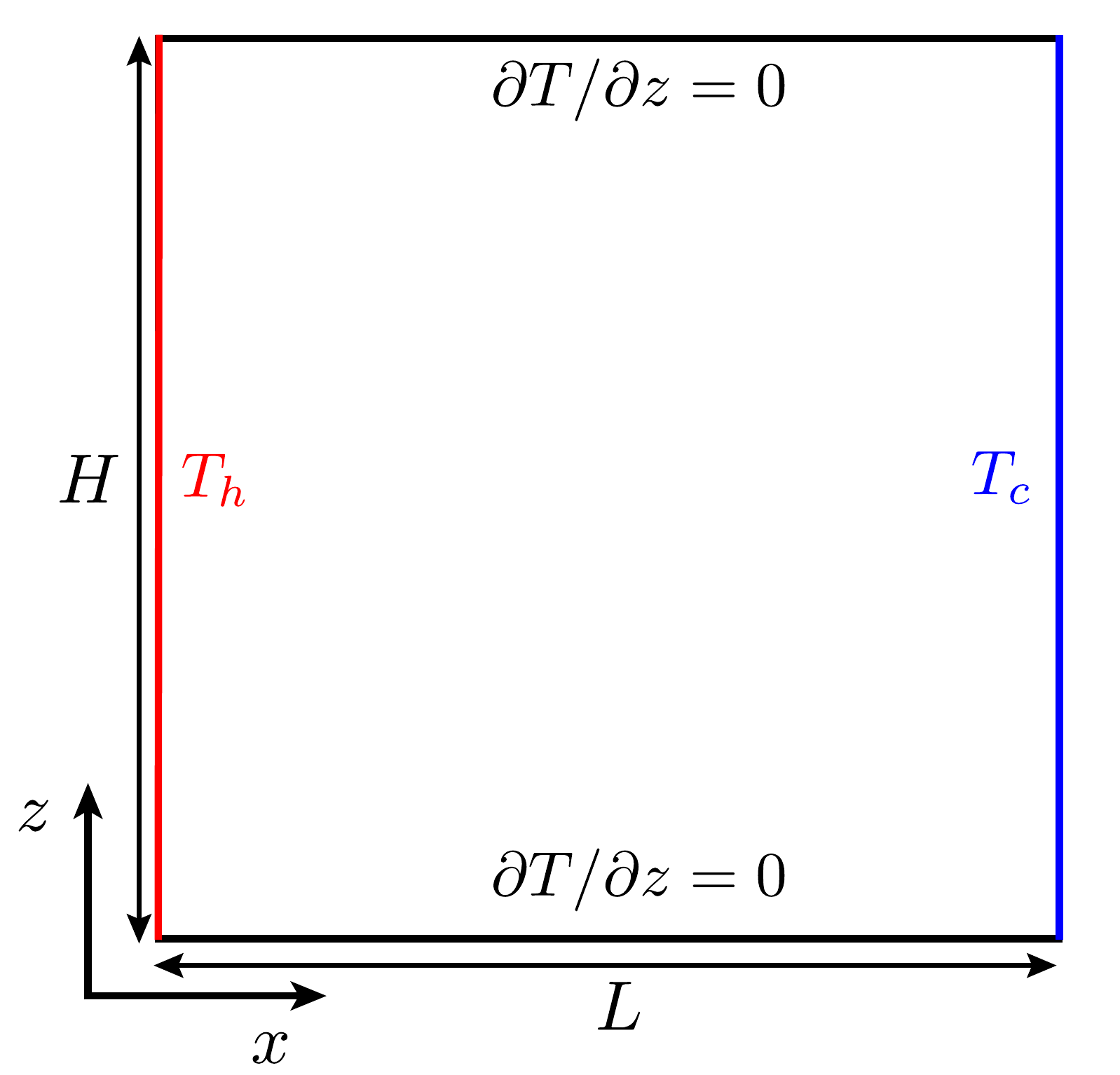}
     \end{overpic}
 \caption{ Sketch of two-dimensional vertical convection with unit aspect ratio. The left vertical wall is heated ($T=T_h$), while the right vertical wall is cooled ($T=T_c$), the temperature difference is $\Delta=T_h-T_c$. The top and bottom walls are adiabatic. All the walls have no-slip and no-penetration velocity boundary conditions.}\label{phase}
\end{figure}

A sketch of 2-D VC is shown in figure \ref{phase}. The top and bottom walls are insulated. The left wall is heated with temperature $T_h$, while the right wall is cooled with temperature $T_c$. No-slip and no-penetration velocity boundary conditions are used at all the walls. The aspect ratio $\Gamma \equiv H/L$ is fixed to 1. The dimensionless governing equations are the incompressible Navier-Stokes equations with Oberbeck–Boussinesq approximation:

\begin{gather}
\nabla\cdot\boldsymbol{u} = 0, \label{eq01}\\
\frac{\partial \boldsymbol{u}}{\partial t} + \boldsymbol{u}\cdot\nabla\boldsymbol{u} = -\nabla p+ \sqrt{\frac{Pr}{Ra}}\nabla^2\boldsymbol{u} + \theta{\vec{\boldsymbol{e}}_z}, \label{eq02}\\
\frac{\partial \theta}{\partial t} + \boldsymbol{u}\cdot\nabla \theta  = \frac{1}{\sqrt{RaPr}}\nabla^2\theta. \label{eq03}
\end{gather}

\noindent  Here $\vec{\boldsymbol{e}}_z$ is the unit vector pointing in the direction opposite to gravity. $\boldsymbol{u}\equiv{(u,w)}$, $\theta$ and $p$ are the dimensionless velocity, temperature and  pressure, respectively. For non-dimensionalization, we use the width of the convection cell $L$ and the free-fall velocity $U={(g\alpha\Delta L)}^{1/2}$. Temperature is nondimensionalized as $\theta=(T-T_c)/\Delta$.

The governing equations were solved using the second-order staggered finite-difference code  AFiD \citep{verzicco1996finite,van2015pencil}. The code has already been extensively used to study RBC \citep{wang2020multiple,wang2020zonal,liu2020two} and internally heated convection \citep{wang2020ihc}. DNSs were performed for $10^7\le Ra\le10^{14}$ with a fixed $Pr=10$. Stretched grids were used to resolve the thin boundary layers and adequate resolutions were ensured to resolve the small scales of turbulence \citep{shishkina2010boundary}. Grids with up to $8192\times8192$ nodes were used for the highest $Ra=10^{14}$. We performed careful grid independence checks for several high-$Ra$ cases. It was found that the difference of $Nu$ and $Re$ for the different grids were always smaller than $1\%$ and $2\%$, respectively. Details on the simulations are provided in the Appendix.

\section{Global flow organization}\label{sec3}

\subsection{Global flow fields}
\begin{figure}
  \centering
  \begin{overpic}[width=1.0\textwidth]{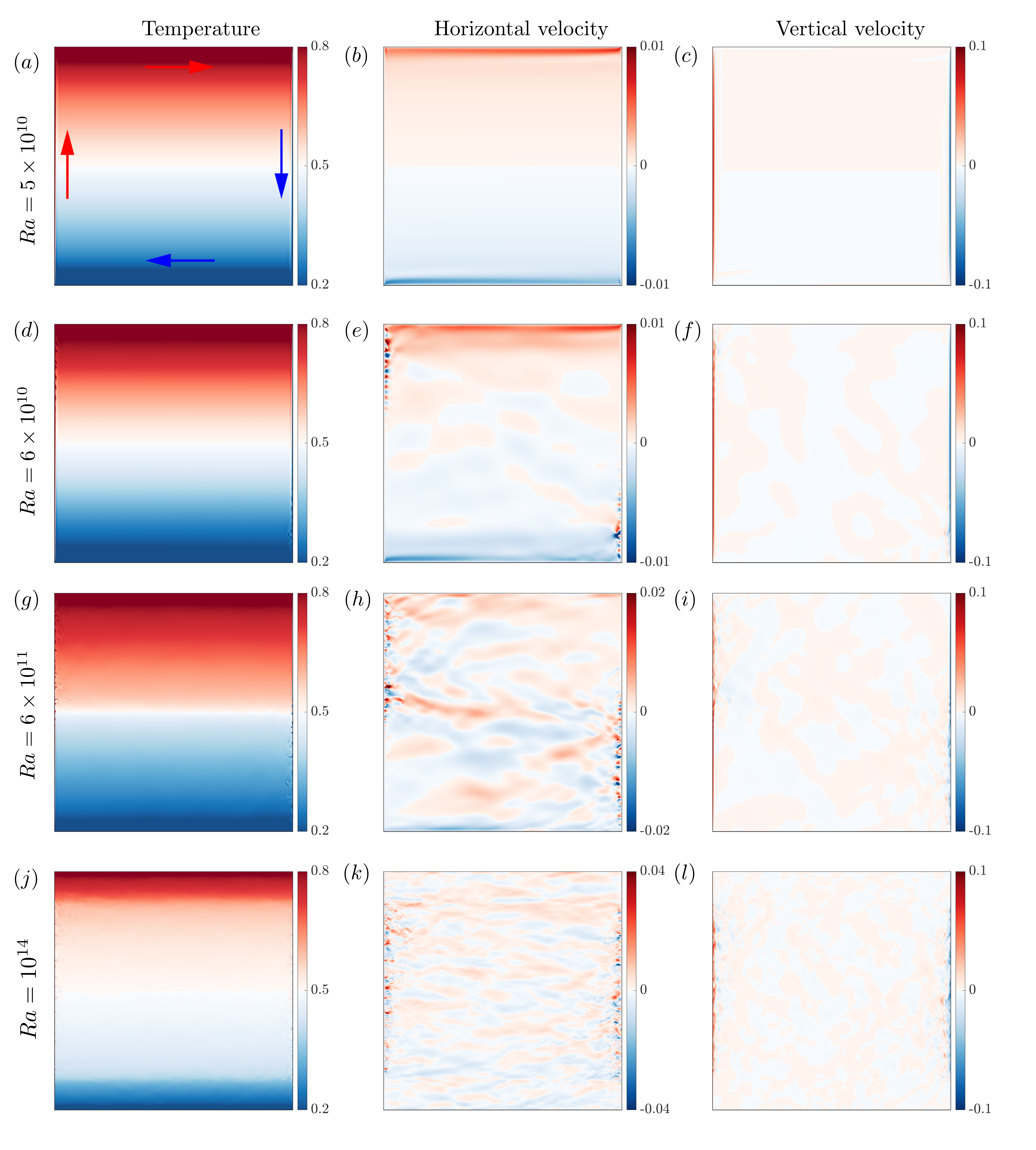}
     \end{overpic}
 \caption{Instantaneous temperature $\theta$ (first column), horizontal velocity $u$ (second column), and vertical velocity (third column) fields for different $Ra$ with $Pr=10$, $\Gamma=1$. (\textit{a-c}) $Ra=5\times 10^{10}$ (regime I). (\textit{d-f}) $Ra=6\times10^{10}$  (regime II). (\textit{g-i}) $Ra=6\times10^{11}$ (regime II). (\textit{j-l}) $Ra=10^{14}$ (regime III). The arrows in panel ($a$) indicate the velocity directions.}\label{flow}
\end{figure}

\noindent We first focus on the change of global flow organizations with increasing $Ra$. Figure \ref{flow} shows instantaneous temperature, horizontal velocity ($u$) and vertical velocity  ($w$) fields for different $Ra$. For the considered $Pr=10$, we find that the flow is still steady for $Ra=5\times10^{10}$ as shown in figures \ref{flow}($a$-$c$), which is consistent with the finding that the critical Rayleigh number $Ra_c$ for the onset of unsteadiness increases with increasing $Pr$ and that the flow is indeed still steady for $Ra=2.5\times10^{10}$ with $Pr=4$ \citep{janssen1995influence}. This is in sharp contrast with RBC where the flow is already turbulent for such high $Ra$ with $Pr=10$ \citep{wang2020multiple}. The flow is stably stratified in the bulk region as shown in figure 
\ref{flow}($a$).The large horizontal velocity regions mainly concentrate near the top and bottom walls (figure \ref{flow}$b$), while the strong vertical motion mainly occurs near the two sidewalls (figure \ref{flow}$c$). Such flow structures are typical for steady VC with large $Pr$ \citep{ravi1994high}.

However, with a minor increase of $Ra$ from $Ra=5\times10^{10}$ to $Ra=6\times10^{10}$, the flow becomes instantaneously chaotic, as shown in figure \ref{flow}($d$-$f$). This finding is consistent with the previous result that for $Pr\ge2.5$, there is immediate transition from the steady to the chaotic flow regime without intermediate regimes \citep{janssen1995influence}. This transition is caused by boundary layer instabilities, which are reflected in the plume ejections in the downstream of the boundary layers (figure \ref{flow}$d$). The strong horizontal/vertical fluid motions still concentrate near the horizontal/vertical walls as indicated in figures \ref{flow}($e$) and \ref{flow}($f$). However, there are already some chaotic features appearing in the bulk, suggesting a change of the bulk properties.

When $Ra$ is further increased to $6\times10^{11}$ (figure \ref{flow}$g$-$i$), further evident changes of the global flow organization appear: (i) The hot plumes mainly eject over the upper half of the hot sidewall, and enter the upper half of the bulk region. This makes the hot upper bulk region more isothermal than the smaller-$Ra$ cases. Similar processes happen for the cold plumes and the lower cold bulk region. Therefore, figure \ref{flow}($g$) clearly shows a larger centre temperature gradient than those in figures \ref{flow}($a$) and \ref{flow}($d$). (ii) The strong horizontal motions now not only occur near the horizontal walls, but also in the bulk region (figure \ref{flow}$h$), and alternating rightward and leftward ``zonal flow'' structures appear.

For the highest $Ra=10^{14}$ (figures \ref{flow}$j$-$l$), the thermal driving is so strong that hot plumes are now ejected over large fractions of the left vertical wall ($0.2 \lesssim z/L \lesssim 1$). The plumes are transported into the bulk region by the zonal flow structures shown in figure \ref{flow}($k$). This process causes efficient mixing in the bulk, which then leads to a smaller centre temperature gradient. Further prominent features are the ``layered'' structures near the top and bottom walls, where relatively hot/cold fluids clearly separate from the near-isothermal bulk region.

\subsection{Mean profiles for temperature and horizontal velocity}

\begin{figure}
  \centering
  \begin{overpic}[width=0.95\textwidth]{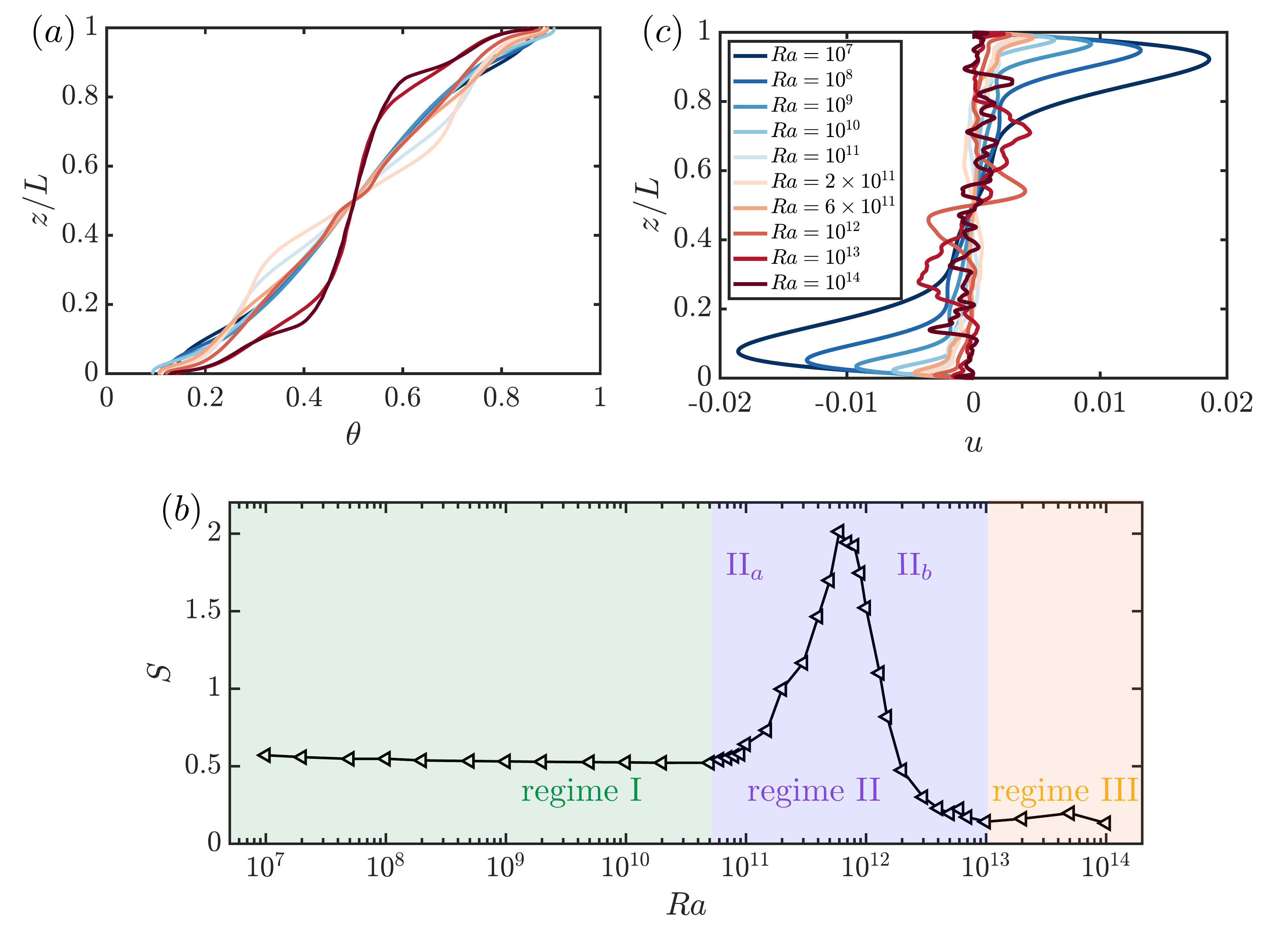}
  \end{overpic}
 \caption{(\textit{a}) Mean temperature profiles $\theta(z)$ at $x/L=0.5$ for different $Ra$ with $Pr=10$. (\textit{b}) Time-averaged centre vertical temperature gradient $S=\left<(L/\Delta)(\partial{T}/\partial{z})_c\right>_t$ as function of $Ra$ for $Pr=10$. In regime I and regime III, $S$ is weakly dependent on $Ra$. In contrast, in regime II, $S$ displays nonmonotonic dependence on $Ra$. regime II is further divided into ${\rm{II}}_a$ and ${\rm{II}}_b$, in which $S$ increases or decreases with increasing $Ra$, respectively. (\textit{c}) Mean horizontal  velocity profiles at $x/L=0.5$ for different $Ra$ with $Pr=10$.  Figure (\textit{a}) and (\textit{c}) share the same legend.}\label{tem_grad}
\end{figure}

We have seen that the global flow organization evidently changes with increasing $Ra$. In this subsection, we quantify these changes by looking at the mean profiles for the temperature and for the horizontal velocity. Figure \ref{tem_grad}($a$) clearly shows the change in the temperature profiles at $x/L=0.5$ with increasing $Ra$, which is consistent with the temperature fields presented in figure \ref{flow}. The change of the bulk temperature profile shape can be quantified by the time-averaged non-dimensional vertical temperature gradient in the cell centre, i.e., $S\equiv\left<(L/\Delta)(\partial{T}/\partial{z})_c\right>_t$ \citep{paolucci1990direct,ravi1994high}. This quantity is plotted in figure \ref{tem_grad}($b$), where one can observe three different regimes. In the well-explored regime I ($Ra \lesssim 5\times10^{10}$), $S$ weakly depends on $Ra$, with a value $S\approx0.5$, close to that of $S=0.52$ for $Ra=10^8$, $Pr=70$ reported in \cite{ravi1994high}. However, in regime II ($5\times10^{10} \lesssim Ra \lesssim 10^{13}$), $S$ has a non-monotonic dependence on $Ra$: it first increases with increasing $Ra$, reaches its maximum at $Ra=6\times10^{11}$, and then decreases again. regime II is further divided into regime ${\rm{II}}_a$, where $S$ increases with increasing $Ra$, and regime ${\rm{II}}_b$, where $S$ decreases with increasing $Ra$. The onset of regime II coincides with the onset of unsteadiness, showing that plume emissions play an important role in altering the bulk properties. The maximum of $S$ happens when about half of sidewall areas (downstream) feature plume emissions, as shown in figure \ref{flow}($g$). Finally, in regime III, $S$ again becomes weakly dependent on $Ra$, while it has a smaller value than that of regime I. The small value of $S$ in regime III is due to the well-mixed bulk region, as can be seen in figure \ref{flow}($j$).

Figure \ref{tem_grad}($c$) shows the change of the horizontal velocity profiles with increasing $Ra$. In regime I, the strong horizontal fluid motions concentrate near the top and bottom walls. In contrast, in regime III,  the largest horizontal velocity appears in the bulk region, and alternating rightward and leftward fluid motions, i.e., zonal flows, are observed even after time average, which is consistent with the instantaneous horizontal velocity filed shown in figure \ref{tem_grad}($k$). Another prominent flow feature of regime III is that the horizontal velocity near the top and bottom walls is close to 0. This means that the ``layered structure'' near the top and bottom walls as indicated in the temperature filed in figure \ref{flow}($j$) is actually nearly a ``dead zone'' with weak fluid motions. Thus the appearance of this nearly ``dead'' layered structure indicates the the onset of regime III. regime II serves to connect regime I and regime III: in regime ${\rm{II}}_a$, the strongest horizontal motion still takes place near the top and bottom walls, see, e.g., the horizontal velocity profile for $Ra=2\times10^{11}$ in figure \ref{tem_grad}($b$). In contrast, in regime ${\rm{II}}_b$, the strongest horizontal fluid motions appear in the bulk, see, e.g., the strong zonal flow motions in the central region for $Ra=10^{12}$, as shown in figure \ref{tem_grad}($b$). We remark that the zonal flow has been found in many geo- and astrophysical flows \citep{yano2003outer,heimpel2005simulation,nadiga2006zonal}, and it has also been extensively studied in RBC \citep{goluskin2014convectively,zhang2020boundary,wang2020zonal,reiter2020generation}. It is remarkable and interesting to also observe zonal flows in the high-$Ra$ VC system. This system thus provides another model to study the physics of the zonal flow.

\subsection{Mean profiles for  temperature and vertical velocity}

\begin{figure}
  \centering
  \begin{overpic}[width=0.95\textwidth]{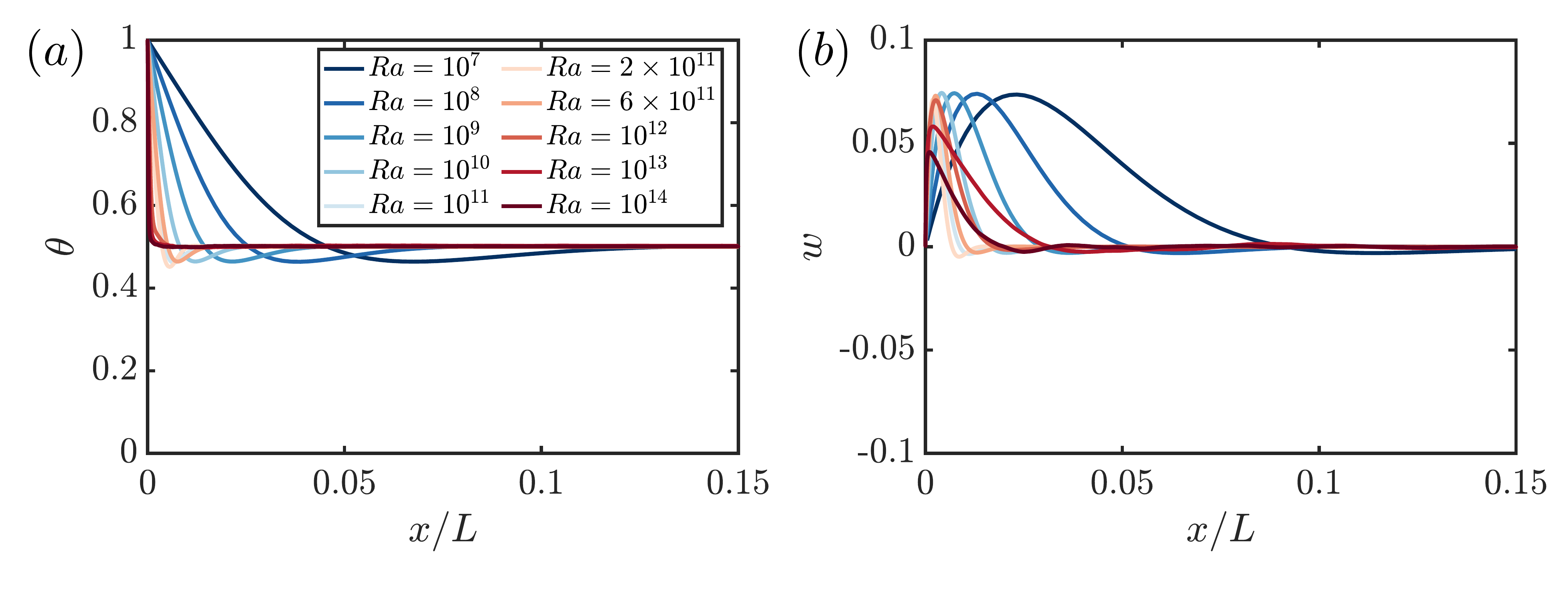}
  \end{overpic}
 \caption{Mean (\textit{a}) longitudinal temperature profile and (\textit{b}) longitudinal profile of the vertical velocity at midheight $z/L=0.5$ for different $Ra$ with $Pr=10$. Panel (\textit{a}) and (\textit{b}) share the same legend.               
 }\label{w_t}
\end{figure}

\noindent We now look at the mean vertical velocity and temperature profiles in longitudinal ($x$) direction. Figure \ref{w_t}($a$) shows the mean temperature profiles n longitudinal at midheight $z/L=0.5$ for different $Ra$. It is seen that for not too high $Ra$, the temperature does not monotonically drop from $\theta=1$ at the hot wall to $\theta=0.5$ in the core. Instead, an undershoot phenomenon is observed. This phenomenon is due to the stable stratification in the bulk \citep{ravi1994high}  and can also be observed in the similarity solutions of the boundary layer equations for natural convection over a vertical hot wall in stably stratified environment \citep{henkes1989laminar}. However, for $Ra\ge10^{12}$, we find that the overshoot  phenomenon disappears. This is due to the continuous emissions of hot plumes at midheight $z/L=0.5$ and beyond, as then the hot fluid directly touches the well-mixed bulk flow with small stratification.

Figure \ref{w_t}($b$) shows vertical velocity profiles in longitudinal direction, again at midheight $z/L=0.5$.  With increasing $Ra$, the boundary layer becomes thiner and the peak vertical velocity becomes smaller. This finding reflects the different flow organizations in the different regimes: the emitted plumes in regimes II and III weaken the overall vertical fluid motions, as compared to the steady flow organization in regime I. It is also reflected in the $Re\sim Ra^\gamma$ scaling, as will be discussed below.


\section{Global heat and momentum transport and dissipation rates} \label{sec4}

\begin{figure}
  \centering
  \vskip 2mm
  \begin{overpic}[width=0.7\textwidth]{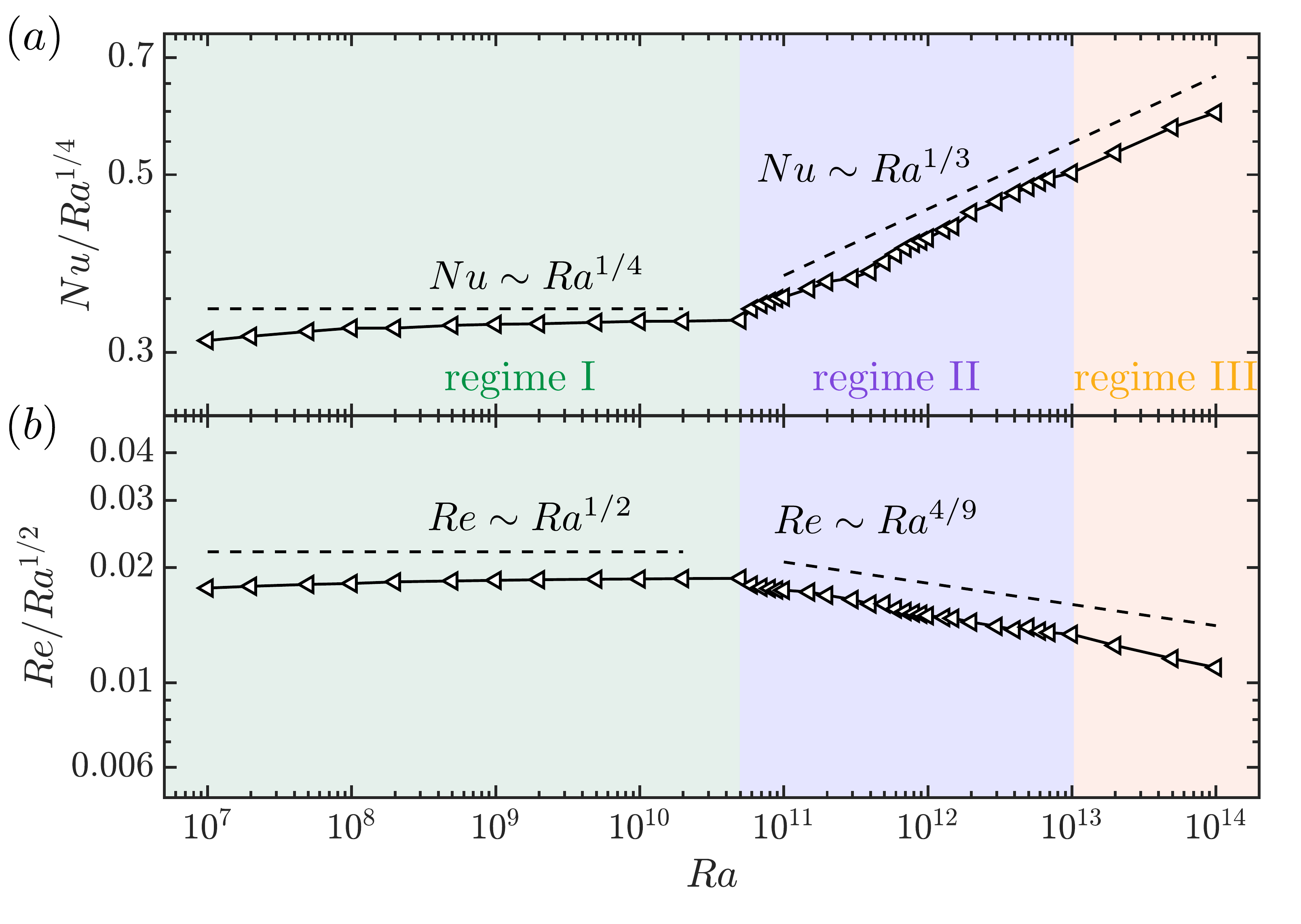}
     \end{overpic}
 \caption{(\textit{a}) Normalized Nusselt number $Nu/Ra^{1/4}$ and (\textit{b}) normalized Reynolds number $Re/Ra^{1/2}$ as functions of $Ra$ for $Pr=10$. The solid lines connect DNS data points, whereas the dashed lines show the suggested scaling laws. There is a clear and sharp transition in scalings between regime I and regime II/III.
 }\label{nu_re}
\end{figure}

\noindent Next, we focus on the global heat ($Nu$) and momentum ($Re$) transport. Here, we use the wind-based Reynolds number with the characteristic velocity 
\begin{gather}
U\equiv \operatorname*{max}\limits_{x}H^{-1}\int_{0}^{H}w dz,
\end{gather}
which is same definition as that in \cite{shishkina2016momentum}.  Figure \ref{nu_re} shows that in regime I, the obtained effective power law scaling relations agree remarkably well with the theoretical predication made for laminar VC \citep{shishkina2016momentum}, namely, $Nu\sim Ra^{1/4}$ and $Re\sim Ra^{1/2}$. The fitted scaling relations are provided in table \ref{tb_scaling}. It is also seen that a slightly faster growth of $Nu$ with $Ra$ is obtained for $Ra\le10^{9}$. A similar increase of the scaling exponent for small $Ra$ was also found before in both confined \citep{shishkina2016momentum,wang2019non} and double periodic VC \citep{ng2015vertical}. However, when $Ra\ge5\times10^{10}$, in regime II and regime III, evidently different scaling relations are observed. The fitted power (see table \ref{tb_scaling} for the obtained values) are pretty close to $Nu\sim Ra^{1/3}$ and $Re\sim Ra^{4/9}$, which, interestingly, were predicted for regime ${\rm{IV}}_u$ of the unifying theory for RBC \citep{grossmann2000scaling}. Such observation of scaling transitions further demonstrates that there are no pure scaling laws in thermal convection. This has already been seen in RBC \citep{grossmann2000scaling,ahlers2009heat}, horizontal convection \citep{shishkina2016heat,shishkina2016prandtl,reiter2020classical}, and internally heat convection \citep{wang2020ihc}, and apparently crossovers between different scaling regimes also occur here. However, the sharpness of the scaling transition from $\beta=1/4$ to $\beta=1/3$ observed here is quite different from the smooth transition seen in RBC. Indeed, in RBC, the transition from $Nu\sim Ra^{1/4}$ to $Nu\sim Ra^{1/3}$ is very smooth, spread over more than two decades in $Ra$ \citep{grossmann2000scaling}, and the linear combination of the $1/4$ and $1/3$ power laws even mimics an effective $2/7$ scaling exponent \citep{castaing1989scaling} over many decades in $Ra$.

\begin{table}
\begin{center}
\begin{tabular}{|c|c|c|c|c|}
\hline
     regime                  &        $Ra$ range                                         & $Nu $                                                & $Re$                                                     & $\left<\epsilon_u\right>/[L^{-4}\nu^{3}(Nu-1)RaPr^{-2}]$  \\
   \hline
     regime I                &      $[10^7,5\times10^{10}]$                        &  $\sim Ra^{0.256}$                          & $\sim Ra^{0.507}$                                           & $\sim Ra^{-0.012}$   \\
     regime II               &      $[5\times10^{10},10^{13}]$                   &  $\sim Ra^{0.330}$                          &  $\sim Ra^{0.438}$                                           &$\sim Ra^{-0.179}$   \\
     regime III              & $[10^{13},10^{14}]$                                   &   $\sim Ra^{0.326}$                         &  $\sim Ra^{0.413}$                                          &  $\sim Ra^{-0.248}$   \\
     \hline
\end{tabular}  
\caption{Fitted scaling relations of $Nu$, $Re$ and $\left<\epsilon_u\right>/[L^{-4}\nu^{3}(Nu-1)RaPr^{-2}]$ with respect to $Ra$ for the three different regimes.}\label{tb_scaling}
\end{center}
\end{table}

\begin{figure}
  \centering
  \vskip 4mm
  \begin{overpic}[width=0.65\textwidth]{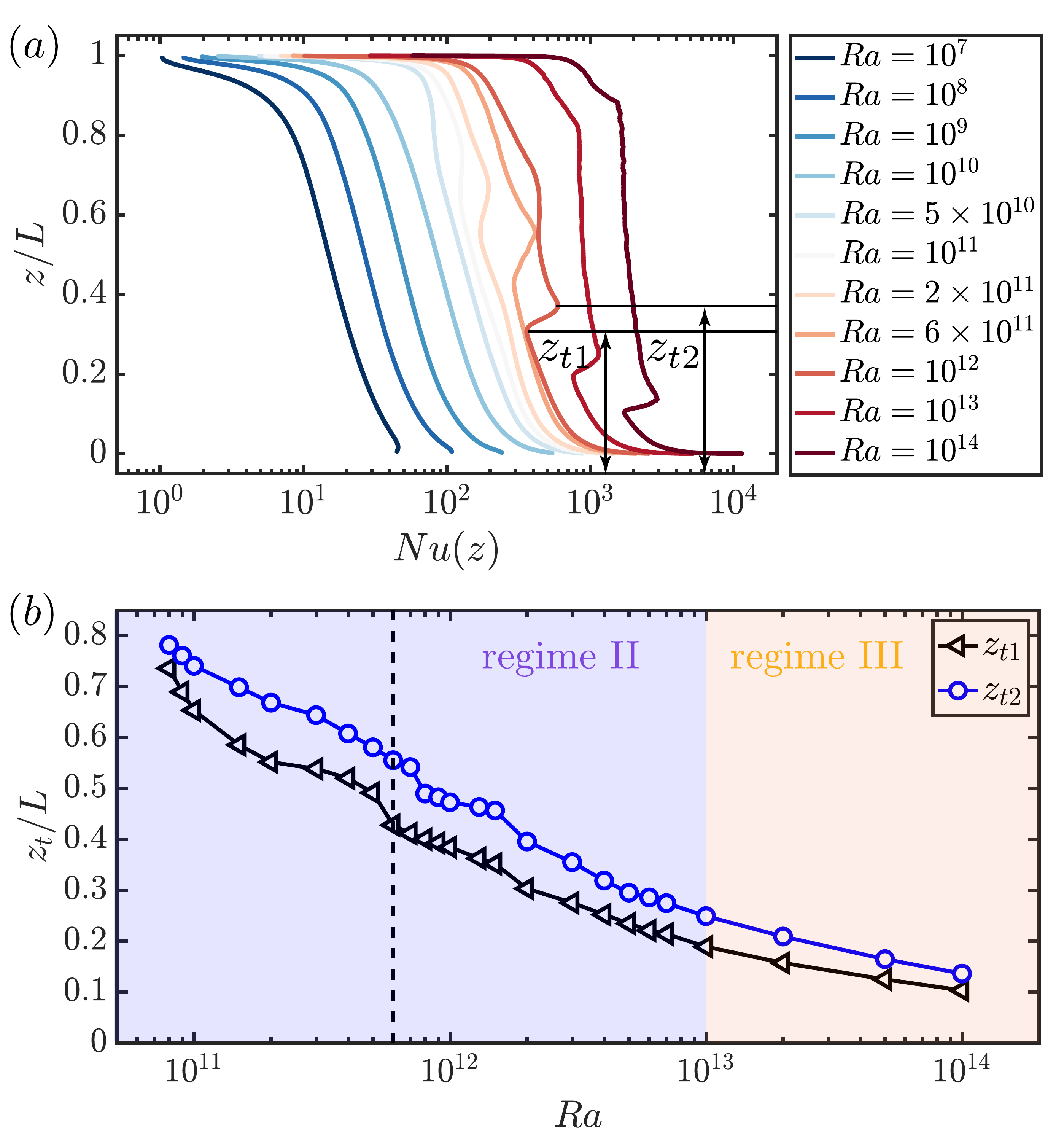}
     \end{overpic}
 \caption{(\textit{a}) Local Nusselt number $Nu(z)$ at the hot wall ($x/L=0$) for different $Ra$, all with $Pr=10$. (\textit{b}) Transition points $z_t/L$ as functions of $Ra$. Here, $z_{t1}$ and $z_{t2}$ are the locations where $Nu(z)$ reaches its local minimum and maximum values, respectively. Such local maximum and minimum occur beyond $Ra\gtrsim 5\times10^{10}$, see panel ($a$). The dashed vertical line denotes the $Ra$ where the centre temperature gradient is maximal.} \label{local}
\end{figure}

In order to better understand the sudden change of the global heat transport properties at thee transition to regime II, we now look at the wall heat flux, which is denoted by the local Nusselt number at the wall  $Nu(z)|_{x=0,1}=\partial \left<\theta\right>_t/\partial x|_{x=0,1}$. Figure \ref{local}($a$) displays $Nu(z)|_{x=0}$ at the left wall, while $Nu(z)|_{x=1}$ at the right wall is not shown due to the inherent symmetry of the system. For $Ra\le5\times10^{10}$, the local $Nu(z)|_{x=0}$ generally decreases monotonically with increasing heights $z$. The large local $Nu(z)|_{x=0}$ for small heights $z$ is attributed to the fact that the hot fluid there is in direct contact with the cold fluid, which leads to large temperature gradients. In contrast, for $Ra>5\times10^{10}$, a local minimum and a local maximum in $Nu(z)|_{x=0}$ are identified, the heights of which are denoted as $z_{t1}$ and $z_{t2}$. It is clearly seen that $Nu(z)|_{x=0}$ after the first transition point $z_{t1}$ increases  compared to the steady cases at $Ra \le 5\times10^{10}$. This is because the emissions of the plumes lead to more efficient shear-driven mixing, and therefore larger local $Nu(z)|_{x=0}$. Thus, the overall heat transport also increases in regime II and later regime III (figure \ref{nu_re}$a$) due the ejections of plumes, and the change of the scaling is also related to the change of the boundary layer properties.

We have shown that the two transition points $z_{t1}$ and $z_{t2}$ roughly correspond to the locations where plumes begin to be ejected. Figure \ref{local}($b$) shows, as expected, that both $z_{t1}$ and $z_{t2}$ decrease with increasing $Ra$, suggesting that the locations where hot plumes begin to be ejected move downwards with increasing $Ra$. At $Ra=6\times10^{11}$, where the centre temperature gradient $S$ achieves its maximum, it can be seen that the mid height $z/L=0.5$ lies between the two transition points, further demonstrating that the maximum of $S$ is achieved once plumes are ejected over approximately half of the area (downstream) of the sidewalls, as seen in the temperature field in figure \ref{flow}($g$).

\begin{figure}
  \centering
  \vskip 2mm
  \begin{overpic}[width=0.8\textwidth]{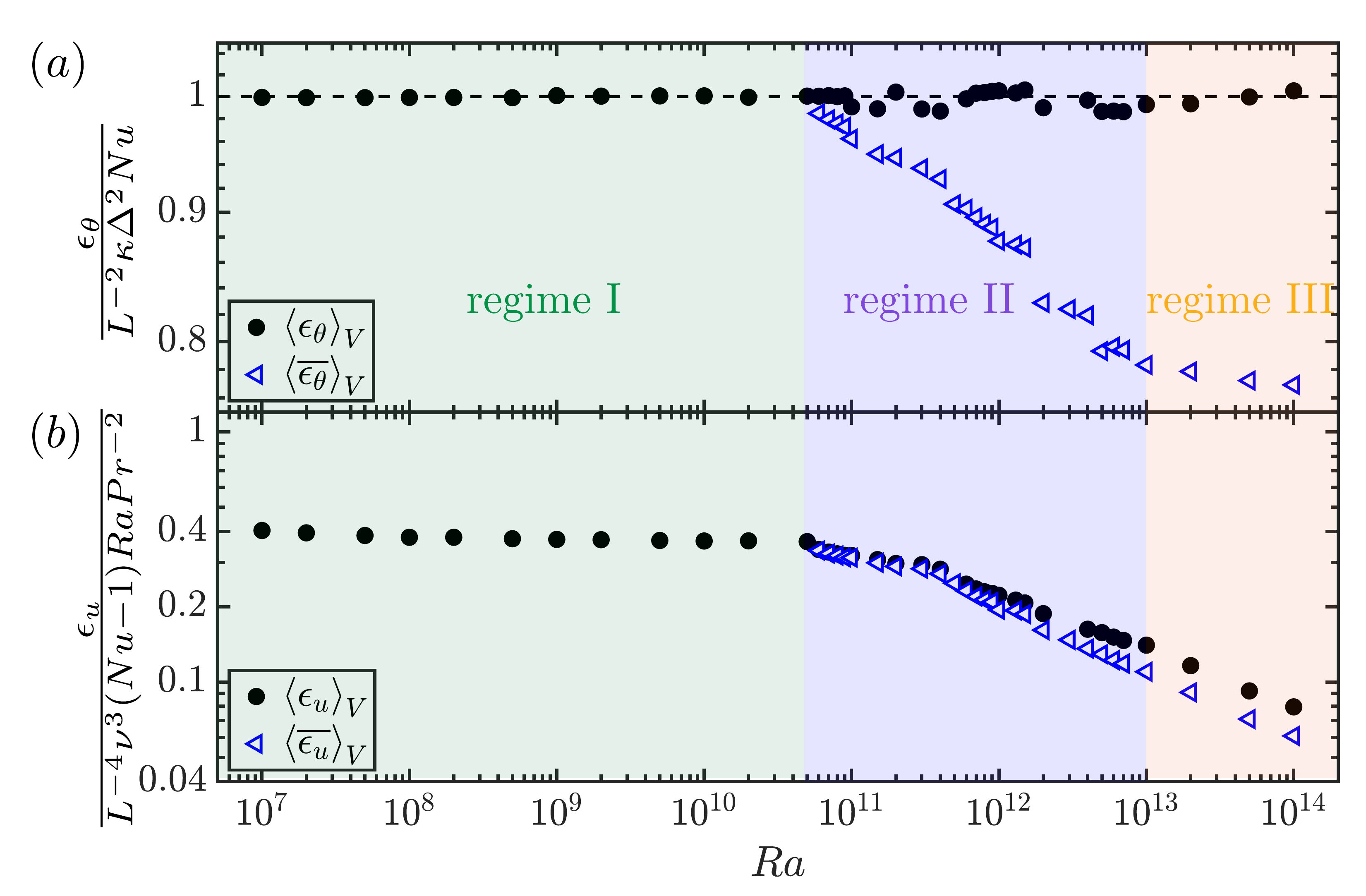}
     \end{overpic}
 \caption{Normalized (\textit{a}) thermal dissipation rate $\left<\epsilon_\theta\right>/(L^{-2}\kappa \Delta^{2} Nu$) and (\textit{b}) kinetic dissipation rate $\left<\epsilon_u\right>/[L^{-4}\nu^{3}(Nu-1)RaPr^{-2}]$ as functions of $Ra$. The black solid circles denote the total dissipation rates while hollow triangles correspond  the dissipation rates of the mean field.}\label{diss}
\end{figure}

Finally, we discuss the thermal and kinetic dissipation rates. In RBC, the following exact relations hold \citep{shraiman1990heat}:
\begin{gather}
\left<\epsilon_u\right>_{V}=\frac{\nu^3}{L^4}(Nu-1)RaPr^{-2},\label{disskinet}\\
\left<\epsilon_\theta\right>_{V}=\kappa\frac{\Delta^2}{L^2}Nu. \label{dissth}
\end{gather}
The average $\left<\right>_V$ is over the whole volume and over time. In VC, relation (\ref{dissth}) still holds, however, relation (\ref{disskinet}) does not hold anymore. Following \cite{ng2015vertical} and \cite{reiter2020classical}, we decompose the dissipation rates into their mean and fluctuating parts: $\left<\epsilon_u\right>_V=\left<\overline{\epsilon_u}\right>_V+\left<\epsilon_u^\prime\right>_V=\nu[\left<(\partial U_i/\partial x_j)^2\right>_V+\left<(\partial u_i^\prime/\partial x_j)^2\right>_V]$. Figure \ref{diss}($a$) shows that relation (\ref{dissth}) is fulfilled in the DNS. It is also seen that the contribution from the mean field decreases with increasing $Ra$, suggesting that with increasing $Ra$, turbulent fluctuations play an increasingly more important role on the mixing process.

The kinetic dissipation rate is displayed in figure \ref{diss}($b$). One sees that the values $\left<\epsilon_u\right>/[L^{-4}\nu^{3}(Nu-1)RaPr^{-2}]$ are always smaller than the corresponding value as occurring in RBC. This was already seen in 3D VC \citep{shishkina2016momentum}. For the steady VC with $Ra\le5\times10^{10}$, the normalized kinetic dissipation rate only weakly depends on $Ra$, as has also been found in 3D VC \citep{shishkina2016momentum}. However, in regimes II and III, it is observed that the normalized kinetic dissipation rate decreases much faster than that in regime I. This can be related to the fact that $Nu$ increases faster in regimes II and III than that in regime I. It is also seen that the contribution from turbulent fluctuations is small, similar as in horizontal convection \citep{reiter2020classical}.

\section{Conclusions} \label{sec5}

In conclusion, we have studied  vertical convection by direct numerical simulations over seven decades of Rayleigh numbers, i.e., $10^7\le Ra\le10^{14}$, for a fixed Prandtl number $Pr=10$ in a two-dimensional convection cell with unit aspect ratio. The main conclusions, corresponding to the answers of the questions put forward in the introduction, are summarized as follows:

\begin{itemize}

 \item [(i)] The dependence of the non-dimensional mean vertical temperature gradient at the cell centre $S$ on $Ra$ shows three different regimes. In regime I ($Ra \lesssim 5\times10^{10}$), $S$ is almost independent of $Ra$, which is consistent with previous work \citep{paolucci1990direct}. However, in the newly identified regime II ($5\times10^{10} \lesssim Ra \lesssim 10^{13}$), $S$ first increases with increasing $Ra$, reaches its maximum, and then decreases again. In regime III ($Ra\gtrsim10^{13}$), $S$ again becomes weakly dependent on $Ra$, with a smaller value than that of regime I. The transition from regime I to regime II coincides with the onset of unsteady fluid motions. The maximum of $S$ occurs when plumes are ejected over about half of the area of the sidewall, namely, in the downstream region. The flow in regime III is characterized by a well-mixed bulk region due to continuous ejection of plumes over large fractions of the sidewalls. Thus $S$ is smaller than that of regime I. 

\item [(ii)]The flow organizations in the three different regimes are quite different from each other. In regime I, the maximal horizontal velocity concentrates near the top and bottom walls. However, the flow gives way to alternating rightward and leftward zonal flows in regime III, where the maximal horizontal velocity appears in the bulk region. Another characteristic feature of the flow in regime III are ``layered'' structures near the top and  bottom walls, where the fluid motions are weak. regime II serves to connect regime I and regime III: in regime  ${\rm{II}}_a$, the maximal velocity still occurs near the top and bottom walls. In contrast, in regime  ${\rm{II}}_b$, the zonal flow structures become more pronounced, and the maximal horizontal velocity is found in the bulk region.

\item[(iii)] Transitions in the scaling relations $Nu\sim Ra^\beta$  and $Re\sim Ra^\gamma$ are found. In regime I, the fitted scaling exponents ($\beta\approx0.26$ and $\gamma\approx0.51$) are in excellent agreement with the theoretical predication of $\beta=1/4$ and $\gamma=1/2$ for the laminar VC \citep{shishkina2016momentum}. However, $\beta$ increases to a value close to 1/3 and $\gamma$ decreases to a value close to 4/9 in regimes II and III. The increased heat transport $Nu$ in regimes II and III is related to the ejection of plumes and larger local heat flux at the sidewalls. The mean kinetic dissipation rate also shows different scalings in the different regimes.

\end{itemize}

We note that the present study only focuses on $Pr=10$. Further studies, both numerical simulations and experiments, are needed to address the influence of the Prandtl number $Pr$ and the aspect ratio $\Gamma$ on the regime transitions in the high-$Ra$ vertical convection. The reported scaling relations for $Nu\sim Ra^\beta$  and $Re\sim Ra^\gamma$ and the observed transitions are however already an important ingredient  to consider to develop a unifying scaling theory over a broad range of control parameters for vertical convection, to finally arrive at the full dependences $Nu(Ra,Pr)$ and $Re(Ra,Pr)$ and their theoretical understanding.

\section*{Acknowledgements}
C.~S.~Ng and K.~L.~Chong are gratefully acknowledged for discussions and support. 
We  also acknowledge the Twente Max-Planck Center,
the Deutsche Forschungsgemeinschaft (Priority Programme SPP 1881 ``Turbulent Superstructures"), PRACE for awarding us access to MareNostrum 4 based in Spain at the Barcelona Computing Center (BSC) under PRACE project 2020225335.
The simulations were partly carried out on the national e-infrastructure of SURFsara, a subsidiary of SURF cooperation, the collaborative ICT organization for Dutch education and research. Q.W. acknowledges financial support from the China Scholarship Council (CSC) and the Natural Science Foundation of China (NSFC) under grant no. 11621202.

\section*{Declaration of interests}
The authors report no conflict of interest.

\section*{Appendix. Tables with simulation details}

\tabcolsep 15pt
\setlength{\LTcapwidth}{0.97\linewidth}
\renewcommand{\arraystretch}{1}
\begin{longtable}{|c|c|c|c|c|c|}

  \hline
       $Ra$                      & $Pr$     & $N_x\times N_z$	           &  $Nu$    &$Re$        & $t_{avg}$           \\
\hline
\endfirsthead
\hline
      $Ra$                      & $Pr$     & $N_x\times N_z$	                  &        $Nu$  &          $Re$      & $t_{avg}$           \\
\hline
\endhead
\hline
\endfoot
              
$10^7$                   &  10     & $256\times256$                      &   17.45     &55.91         &    s                                       \\
$2\times10^7$       &  10     & $256\times256$                      &   21.01    & 79.94          &    s                                \\
$5\times10^5$       &  10     & $256\times256$                      &  26.79       &127.81       &    s                                     \\
$10^8$                   &  10     & $256\times256$                      &   32.17      &181.76       &    s                                  \\
$2\times10^8$        &  10     & $512\times512$                      &  38.25       &259.57      &    s                                       \\
$5\times10^8$        &  10     & $512\times512$                      &   48.48      &412.66       &   s                                   \\
$10^9$                   &  10     & $512\times512$                      &  57.83       &585.72      &   s                                 \\
$2\times10^9$       &  10     & $1024\times1024$                      &   68.87    &831.61          &   s                                    \\
$5\times10^9$       &  10     & $1024\times1024$                      &   86.97    &1305.75        &    s                                     \\
$\textcolor{blue}{\it{10^{10}}}$               &  \textcolor{blue}{\it{10}}     & $\textcolor{blue}{\it{512\times512}}$                     &   \textcolor{blue}{\it{104.10}}        &   \textcolor{blue}{\it{1864.73}}      & \textcolor{blue}{\it{s}}        \\
$10^{10}$              &  10     & $1024\times1024$                      & 103.72     &1867.66        &   s                                   \\
$\textcolor{blue}{\it{10^{10}}}$               &  \textcolor{blue}{\it{10}}     & $\textcolor{blue}{\it{2048\times2048}}$                     &   \textcolor{blue}{\it{103.57}}        &   \textcolor{blue}{\it{1869.23}}      & \textcolor{blue}{\it{s}}        \\
$2\times10^{10}$   &  10     & $2048\times2048$                      &   123.38     &2647.58       & s                                         \\
$5\times10^{10}$   &  10     & $2048\times2048$                      &  155.63      &4191.51      & s                                \\
$6\times10^{10}$   &  10     & $2048\times2048$                      &  168.23        & 4411.20      &   4000                               \\
$7\times10^{10}$   &  10     & $2048\times2048$                      &  177.07        & 4698.16      & 4000                                 \\
$8\times10^{10}$   &  10     & $2048\times2048$                      &   184.72       &  4989.32     &   4000                               \\
$9\times10^{10}$   &  10     & $2048\times2048$                      &   191.77       &   5243.58    &    4000                              \\
$10^{11}$               &  10     & $2048\times2048$                      &  197.76      &5519.30     &  700                                         \\
$1.5\times10^{11}$   &  10     & $2048\times2048$                      &  224.16        &  6688.48     & 600                                 \\
$2\times10^{11}$    &  10     & $2048\times2048$                      &  245.75     & 7571.25          & 1000                                         \\
$3\times10^{11}$   &  10     & $4096\times4096$                      &   274.91       &  9035.54     &  600                                \\
$4\times10^{11}$   &  10     & $4096\times4096$                      &   301.02       &  10157.67     &   600                               \\
$5\times10^{11}$    &  10     & $4096\times4096$                      &   327.13      & 11371.01     & 400                                          \\
$6\times10^{11}$   &  10     & $4096\times4096$                      &  350.12        & 12071.13      &   400                               \\
$7\times10^{11}$   &  10     & $4096\times4096$                      &  370.41        & 12852.20      &    600                              \\
$8\times10^{11}$   &  10     & $4096\times4096$                      &  388.00        &   13599.04    & 500                                 \\
$9\times10^{11}$   &  10     & $4096\times4096$                      &  402.48        &   14329.93    &  500                                \\
$10^{12}$               &  10     & $4096\times4096$                      &   416.76     &    14969.80     & 538                                       \\
$1.3\times10^{12}$   &  10     & $4096\times4096$                      &  455.39        & 16886.46      &  318                                \\
$1.5\times10^{12}$   &  10     & $4096\times4096$                      &  476.91        & 18047.30      &   304                               \\
$2\times10^{12}$    &  10     & $4096\times4096$                      &  533.62      &20368.53   &  500                                            \\
$3\times10^{12}$   &  10     & $4096\times4096$                      &  608.89        & 24315.69       &  307                                \\
$4\times10^{12}$   &  10     & $4096\times4096$                      &   670.80       &  27518.31     &   600                               \\
$5\times10^{12}$    &  10     & $4096\times4096$                      &  721.14   & 31241.75        & 400                                         \\
$\textcolor{blue}{\it{5\times10^{12}}}$               &  \textcolor{blue}{\it{10}}     & $\textcolor{blue}{\it{6144\times6144}} $                   &   \textcolor{blue}{\it{719.14}}     &    \textcolor{blue}{\it{30901.58}}     & \textcolor{blue}{\it{200}}        \\
$6\times10^{12}$    &  10     & $4096\times4096$                      &  766.01  &  33418.22      & 339                                       \\
$7\times10^{12}$    &  10     & $4096\times4096$                      & 804.78  &  35765.90      &   442                                     \\
$10^{13}$               &  10    & $6144\times6144$                      &  894.12    & 42328.91     & 170                                 \\
$\textcolor{blue}{\it{10^{13}}}$               &  \textcolor{blue}{\it{10}}     & $\textcolor{blue}{\it{4096\times4096}}$                     &   \textcolor{blue}{\it{896.63}}      &   \textcolor{blue}{\it{43023.82}}      & \textcolor{blue}{\it{400}}        \\
$2\times10^{13}$               &  10    & $6144\times6144$                      &  1125.87      & 55947.28   & 250                                 \\
$5\times10^{13}$               &  10    & $6144\times6144$                      &  1523.18     &81719.17 & 200                                 \\
$10^{14}$                          &  10    & $8192\times8192$                      &   1890.36     & 109546.69   & 160                                 \\
$\textcolor{blue}{\it{10^{14}}}$               &  \textcolor{blue}{\it{10}}     & $\textcolor{blue}{\it{6144\times6144}}$                     &   $\textcolor{blue}{\it{1907.95}} $     & $\textcolor{blue}{\it{109022.39}}$        & \textcolor{blue}{\it{150}}        \\
\hline
 \caption{The columns from left to right indicate the following: the Rayleigh number $Ra$, the Prandtl number $Pr$, the grid resolution $N_x\times N_z$, the Nusselt number $Nu$, the Reynolds number $Re$,  the time $t_{avg}$ used to average $Nu$ and $Re$. The aspect ratio is fixed to 1 for all the cases. ``s'' means that the flow is steady. Cases indicated in blue and italic are used for grid independence checks. We note that the difference of $Nu$ for two different grids is always smaller than $1\%$, and the difference of $Re$ for the different grids is always smaller than $2\%$.}\label{tabs}
\end{longtable}

\bibliographystyle{jfm}
\bibliography{2d_vc}

\begin{thebibliography}{60}
\expandafter\ifx\csname natexlab\endcsname\relax\def\natexlab#1{#1}\fi
\def\au#1{#1} \def\ed#1{#1} \def\yr#1{#1}\def\at#1{#1}\def\jt#1{\textit{#1}}
  \def\bt#1{#1}\def\bvol#1{\textbf{#1}} \def\vol#1{#1} \def\pg#1{#1}
  \def\publ#1{#1}\def\arxiv#1{#1}\def\org#1{#1}\def\st#1{\textit{#1}}

\bibitem[Ahlers {\em et~al.\/}(2009)Ahlers, Grossmann \& Lohse]{ahlers2009heat}
{\sc \au{Ahlers, G.}, \au{Grossmann, S.} \& \au{Lohse, D.}} \yr{2009}
  \at{{Heat transfer and large scale dynamics in turbulent Rayleigh-B{\'e}nard
  convection}}.  \jt{Rev. Mod. Phys.}  \bvol{81},  \pg{503--537}.

\bibitem[Batchelor(1954)]{batchelor1954heat}
{\sc \au{Batchelor, G.~K.}} \yr{1954}  \at{{Heat transfer by free convection
  across a closed cavity between vertical boundaries at different
  temperatures}}.  \jt{Quart. Appl. Math.}  \bvol{12},  \pg{209--233}.

\bibitem[Blythe {\em et~al.\/}(1983)Blythe, Daniels \&
  Simpkins]{blythe1983thermal}
{\sc \au{Blythe, P.~A.}, \au{Daniels, P.~G.} \& \au{Simpkins, P.~G.}} \yr{1983}
   \at{Thermal convection in a cavity: the core structure near the horizontal
  boundaries}.  \jt{Proc. R. Soc. Lond. A}  \bvol{387},  \pg{367--388}.

\bibitem[Castaing {\em et~al.\/}(1989)Castaing, Gunaratne, Heslot, Kadanoff,
  Libchaber, Thomae, Wu, Zaleski \& Zanetti]{castaing1989scaling}
{\sc \au{Castaing, B.}, \au{Gunaratne, G.}, \au{Heslot, F.}, \au{Kadanoff, L.},
  \au{Libchaber, A.}, \au{Thomae, S.}, \au{Wu, X.-Z.}, \au{Zaleski, S.} \&
  \au{Zanetti, G.}} \yr{1989}  \at{{Scaling of hard thermal turbulence in
  Rayleigh-B{\'e}nard convection}}.  \jt{J. Fluid Mech.}  \bvol{204},
  \pg{1--30}.

\bibitem[Chenoweth \& Paolucci(1986)]{chenoweth1986natural}
{\sc \au{Chenoweth, D.~R.} \& \au{Paolucci, S}} \yr{1986}  \at{{Natural
  convection in an enclosed vertical air layer with large horizontal
  temperature differences}}.  \jt{J. Fluid Mech.}  \bvol{169},  \pg{173--210}.

\bibitem[Chill{\`a} \& Schumacher(2012)]{chilla2012new}
{\sc \au{Chill{\`a}, F.} \& \au{Schumacher, J.}} \yr{2012}  \at{{New
  perspectives in turbulent Rayleigh-B{\'e}nard convection}}.  \jt{Eur. Phys.
  J. E}  \bvol{35},  \pg{58}.

\bibitem[Chong {\em et~al.\/}(2020)Chong, Yang, Wang, Verzicco \&
  Lohse]{chong2020caf}
{\sc \au{Chong, K.~L.}, \au{Yang, R.}, \au{Wang, Q.}, \au{Verzicco, R.} \&
  \au{Lohse, D.}} \yr{2020}  \at{{Caf{\'e} Latte: Spontaneous layer formation
  in laterally cooled double diffusive convection}}.  \jt{J. Fluid Mech.}
  \bvol{900},  \pg{R6}.

\bibitem[Gill(1966)]{gill1966boundary}
{\sc \au{Gill, A.~E.}} \yr{1966}  \at{The boundary-layer regime for convection
  in a rectangular cavity}.  \jt{J. Fluid Mech.}  \bvol{26},  \pg{515--536}.

\bibitem[Goluskin {\em et~al.\/}(2014)Goluskin, Johnston, Flierl \&
  Spiegel]{goluskin2014convectively}
{\sc \au{Goluskin, D.}, \au{Johnston, H.}, \au{Flierl, G.~R.} \& \au{Spiegel,
  E.~A.}} \yr{2014}  \at{Convectively driven shear and decreased heat flux}.
  \jt{J. Fluid Mech.}  \bvol{759},  \pg{360--385}.

\bibitem[Graebel(1981)]{graebel1981influence}
{\sc \au{Graebel, W.~P.}} \yr{1981}  \at{{The influence of Prandtl number on
  free convection in a rectangular cavity}}.  \jt{Int. J. Heat Mass Transf.}
  \bvol{24},  \pg{125--131}.

\bibitem[Grossmann \& Lohse(2000)]{grossmann2000scaling}
{\sc \au{Grossmann, S.} \& \au{Lohse, D.}} \yr{2000}  \at{Scaling in thermal
  convection: a unifying theory}.  \jt{J. Fluid Mech.}  \bvol{407},
  \pg{27--56}.

\bibitem[Grossmann \& Lohse(2001)]{grossmann2001thermal}
{\sc \au{Grossmann, S.} \& \au{Lohse, D.}} \yr{2001}  \at{Thermal convection
  for large prandtl numbers}.  \jt{Phys. Rev. Lett.}  \bvol{86},  \pg{3316}.

\bibitem[Grossmann \& Lohse(2002)]{grossmann2002prandtl}
{\sc \au{Grossmann, S.} \& \au{Lohse, D.}} \yr{2002}  \at{{Prandtl and Rayleigh
  number dependence of the Reynolds number in turbulent thermal convection}}.
  \jt{Phys. Rev. E}  \bvol{66},  \pg{016305}.

\bibitem[Grossmann \& Lohse(2004)]{grossmann2004fluctuations}
{\sc \au{Grossmann, S.} \& \au{Lohse, D.}} \yr{2004}  \at{{Fluctuations in
  turbulent Rayleigh--B{\'e}nard convection: the role of plumes}}.  \jt{Phys.
  Fluids}  \bvol{16},  \pg{4462--4472}.

\bibitem[Guo {\em et~al.\/}(2015)Guo, Zhou, Cen, Qu, Lu, Sun \&
  Shang]{guo2015effect}
{\sc \au{Guo, S.-X.}, \au{Zhou, S.-Q.}, \au{Cen, X.-R.}, \au{Qu, L.}, \au{Lu,
  Y.-Z.}, \au{Sun, L.} \& \au{Shang, X.-D.}} \yr{2015}  \at{{The effect of cell
  tilting on turbulent thermal convection in a rectangular cell}}.  \jt{J.
  Fluid Mech.}  \bvol{762},  \pg{273--287}.

\bibitem[Hadley(1735)]{hadley1735vi}
{\sc \au{Hadley, George}} \yr{1735}  \at{Concerning the cause of the general
  trade-winds}.  \jt{Philos. Trans. R. Soc. London.}  \bvol{39},  \pg{58--62}.

\bibitem[Heimpel {\em et~al.\/}(2005)Heimpel, Aurnou \&
  Wicht]{heimpel2005simulation}
{\sc \au{Heimpel, M.}, \au{Aurnou, J.} \& \au{Wicht, J.}} \yr{2005}
  \at{{Simulation of equatorial and high-latitude jets on Jupiter in a deep
  convection model}}.  \jt{Nature}  \bvol{438},  \pg{193--196}.

\bibitem[Henkes \& Hoogendoorn(1989)]{henkes1989laminar}
{\sc \au{Henkes, R. A. W.~M.} \& \au{Hoogendoorn, C.~J.}} \yr{1989}
  \at{Laminar natural convection boundary-layer flow along a heated vertical
  plate in a stratified environment}.  \jt{Int. J. Heat Mass Transf.}
  \bvol{32},  \pg{147--155}.

\bibitem[Janssen \& Henkes(1995)]{janssen1995influence}
{\sc \au{Janssen, R. J.~A.} \& \au{Henkes, R. A. W.~M.}} \yr{1995}
  \at{Influence of prandtl number on instability mechanisms and transition in a
  differentially heated square cavity}.  \jt{J. Fluid Mech.}  \bvol{290},
  \pg{319--344}.

\bibitem[Lappa(2009)]{lappa2009thermal}
{\sc \au{Lappa, M.}} \yr{2009} {\em Thermal convection: patterns, evolution and
  stability\/}.  \publ{John Wiley \& Sons}.

\bibitem[Le~Qu{\'e}r{\'e} \& Behnia(1998)]{le1998onset}
{\sc \au{Le~Qu{\'e}r{\'e}, P.} \& \au{Behnia, M.}} \yr{1998}  \at{{From onset
  of unsteadiness to chaos in a differentially heated square cavity}}.  \jt{J.
  Fluid Mech.}  \bvol{359},  \pg{81--107}.

\bibitem[Liu {\em et~al.\/}(2020)Liu, Chong, Wang, Ng, Verzicco \&
  Lohse]{liu2020two}
{\sc \au{Liu, H.-R.}, \au{Chong, K.~L.}, \au{Wang, Q.}, \au{Ng, C.~S.},
  \au{Verzicco, R.} \& \au{Lohse, D.}} \yr{2020}  \at{Two-layer thermally
  driven turbulence: Mechanisms for interface breakup}.  \jt{arXiv:2005.05633}
  .

\bibitem[Lohse \& Xia(2010)]{lohse2010small}
{\sc \au{Lohse, D.} \& \au{Xia, K.-Q.}} \yr{2010}  \at{{Small-scale properties
  of turbulent Rayleigh-B{\'e}nard convection}}.  \jt{Annu. Rev. Fluid Mech}
  \bvol{42},  \pg{335--364}.

\bibitem[Nadiga(2006)]{nadiga2006zonal}
{\sc \au{Nadiga, B.~T.}} \yr{2006}  \at{{On zonal jets in oceans}}.
  \jt{Geophys. Res. Lett.}  \bvol{33},  \pg{L10601}.

\bibitem[Ng {\em et~al.\/}(2015)Ng, Ooi, Lohse \& Chung]{ng2015vertical}
{\sc \au{Ng, C.~S.}, \au{Ooi, A.}, \au{Lohse, D.} \& \au{Chung, D.}} \yr{2015}
  \at{{Vertical natural convection: application of the unifying theory of
  thermal convection}}.  \jt{J. Fluid Mech.}  \bvol{764},  \pg{349--361}.

\bibitem[Ng {\em et~al.\/}(2017)Ng, Ooi, Lohse \& Chung]{ng2017changes}
{\sc \au{Ng, C.~S.}, \au{Ooi, A.}, \au{Lohse, D.} \& \au{Chung, D.}} \yr{2017}
  \at{{Changes in the boundary-layer structure at the edge of the ultimate
  regime in vertical natural convection}}.  \jt{J. Fluid Mech.}  \bvol{825},
  \pg{550--572}.

\bibitem[Ng {\em et~al.\/}(2018)Ng, Ooi, Lohse \& Chung]{ngbulk}
{\sc \au{Ng, C.~S.}, \au{Ooi, A.}, \au{Lohse, D.} \& \au{Chung, D.}} \yr{2018}
  \at{Bulk scaling in wall-bounded and homogeneous vertical natural
  convection}.  \jt{J. Fluid Mech.}  \bvol{841},  \pg{825--850}.

\bibitem[Ng {\em et~al.\/}(2020)Ng, Spandan, Verzicco \& Lohse]{ng2020non}
{\sc \au{Ng, C.~S.}, \au{Spandan, V.}, \au{Verzicco, R.} \& \au{Lohse, D.}}
  \yr{2020}  \at{Non-monotonic transport mechanisms in vertical natural
  convection with dispersed light droplets}.  \jt{J. Fluid Mech.}  \bvol{900},
  \pg{A34}.

\bibitem[Paolucci(1990)]{paolucci1990direct}
{\sc \au{Paolucci, S.}} \yr{1990}  \at{Direct numerical simulation of
  two-dimensional turbulent natural convection in an enclosed cavity}.  \jt{J.
  Fluid Mech.}  \bvol{215},  \pg{229--262}.

\bibitem[Paolucci \& Chenoweth(1989)]{paolucci1989transition}
{\sc \au{Paolucci, S.} \& \au{Chenoweth, D.~R.}} \yr{1989}  \at{Transition to
  chaos in a differentially heated vertical cavity}.  \jt{J. Fluid Mech.}
  \bvol{201},  \pg{379--410}.

\bibitem[van~der Poel {\em et~al.\/}(2015)van~der Poel, Ostilla-M{\'o}nico,
  Donners \& Verzicco]{van2015pencil}
{\sc \au{van~der Poel, E.~P.}, \au{Ostilla-M{\'o}nico, R.}, \au{Donners, J.} \&
  \au{Verzicco, R.}} \yr{2015}  \at{{A pencil distributed finite difference
  code for strongly turbulent wall-bounded flows}}.  \jt{Comput.Fluids}
  \bvol{116},  \pg{10--16}.

\bibitem[van~der Poel {\em et~al.\/}(2013)van~der Poel, Stevens \&
  Lohse]{van2013comparison}
{\sc \au{van~der Poel, E.~P.}, \au{Stevens, R. J. A.~M.} \& \au{Lohse, D.}}
  \yr{2013}  \at{{Comparison between two-and three-dimensional
  Rayleigh--B{\'e}nard convection}}.  \jt{J. Fluid Mech.}  \bvol{736},
  \pg{177--194}.

\bibitem[Ravi {\em et~al.\/}(1994)Ravi, Henkes \& Hoogendoorn]{ravi1994high}
{\sc \au{Ravi, M.~R.}, \au{Henkes, R. A. W.~M.} \& \au{Hoogendoorn, C.~J.}}
  \yr{1994}  \at{{On the high-Rayleigh-number structure of steady laminar
  natural-convection flow in a square enclosure}}.  \jt{J. Fluid Mech.}
  \bvol{262},  \pg{325--351}.

\bibitem[Reiter \& Shishkina(2020)]{reiter2020classical}
{\sc \au{Reiter, P.} \& \au{Shishkina, O.}} \yr{2020}  \at{Classical and
  symmetrical horizontal convection: detaching plumes and oscillations}.
  \jt{J. Fluid Mech.}  \bvol{892},  \pg{R1}.

\bibitem[Reiter {\em et~al.\/}(2020)Reiter, Zhang, Stepanov \&
  Shishkina]{reiter2020generation}
{\sc \au{Reiter, P.}, \au{Zhang, X.}, \au{Stepanov, R.} \& \au{Shishkina, O.}}
  \yr{2020}  \at{Generation of zonal flows in convective systems by travelling
  thermal waves}.  \jt{arXiv:2009.01735} .

\bibitem[Shishkina(2016)]{shishkina2016momentum}
{\sc \au{Shishkina, O.}} \yr{2016}  \at{Momentum and heat transport scalings in
  laminar vertical convection}.  \jt{Phys. Rev. E}  \bvol{93},  \pg{051102}.

\bibitem[Shishkina {\em et~al.\/}(2016)Shishkina, Grossmann \&
  Lohse]{shishkina2016heat}
{\sc \au{Shishkina, O.}, \au{Grossmann, S.} \& \au{Lohse, D.}} \yr{2016}
  \at{Heat and momentum transport scalings in horizontal convection}.
  \jt{Geophys. Res. Lett.}  \bvol{43},  \pg{1219--1225}.

\bibitem[Shishkina \& Horn(2016)]{shishkina2016thermal}
{\sc \au{Shishkina, O.} \& \au{Horn, S.}} \yr{2016}  \at{Thermal convection in
  inclined cylindrical containers}.  \jt{J. Fluid Mech.}  \bvol{790},  \pg{R3}.

\bibitem[Shishkina {\em et~al.\/}(2010)Shishkina, Stevens, Grossmann \&
  Lohse]{shishkina2010boundary}
{\sc \au{Shishkina, O.}, \au{Stevens, R. J. A.~M.}, \au{Grossmann, S.} \&
  \au{Lohse, D.}} \yr{2010}  \at{{Boundary layer structure in turbulent thermal
  convection and its consequences for the required numerical resolution}}.
  \jt{New J. Phys.}  \bvol{12},  \pg{075022}.

\bibitem[Shishkina \& Wagner(2016)]{shishkina2016prandtl}
{\sc \au{Shishkina, O.} \& \au{Wagner, S.}} \yr{2016}  \at{{Prandtl-number
  dependence of heat transport in laminar horizontal convection}}.  \jt{Phys.
  Rev. Lett.}  \bvol{116},  \pg{024302}.

\bibitem[Shraiman \& Siggia(1990)]{shraiman1990heat}
{\sc \au{Shraiman, B.~I.} \& \au{Siggia, E.~D.}} \yr{1990}  \at{{Heat transport
  in high-Rayleigh-number convection}}.  \jt{Phys. Rev. A}  \bvol{42},
  \pg{3650}.

\bibitem[Stevens {\em et~al.\/}(2013)Stevens, van~der Poel, Grossmann \&
  Lohse]{stevens2013unifying}
{\sc \au{Stevens, R. J. A.~M.}, \au{van~der Poel, E.~P.}, \au{Grossmann, S.} \&
  \au{Lohse, D.}} \yr{2013}  \at{The unifying theory of scaling in thermal
  convection: the updated prefactors}.  \jt{J. Fluid Mech.}  \bvol{730},
  \pg{295--308}.

\bibitem[Tanny \& Tsinober(1988)]{tanny1988dynamics}
{\sc \au{Tanny, J.} \& \au{Tsinober, A.~B.}} \yr{1988}  \at{The dynamics and
  structure of double-diffusive layers in sidewall-heating experiments}.
  \jt{J. Fluid Mech.}  \bvol{196},  \pg{135--156}.

\bibitem[Thorpe {\em et~al.\/}(1969)Thorpe, Hutt \& Soulsby]{thorpe1969effect}
{\sc \au{Thorpe, S.~A.}, \au{Hutt, P.~K.} \& \au{Soulsby, R.}} \yr{1969}
  \at{The effect of horizontal gradients on thermohaline convection}.  \jt{J.
  Fluid Mech.}  \bvol{38}~(2),  \pg{375--400}.

\bibitem[Trias {\em et~al.\/}(2010)Trias, Gorobets, Soria \&
  Oliva]{trias2010direct}
{\sc \au{Trias, F.~X.}, \au{Gorobets, A.}, \au{Soria, M.} \& \au{Oliva, A.}}
  \yr{2010}  \at{{Direct numerical simulation of a differentially heated cavity
  of aspect ratio 4 with Rayleigh numbers up to $10^{11}$--Part I: Numerical
  methods and time-averaged flow}}.  \jt{Int. J. Heat Mass Transf.}  \bvol{53},
   \pg{665--673}.

\bibitem[Trias {\em et~al.\/}(2007)Trias, Soria, Oliva \&
  P{\'e}rez-Segarra]{trias2007direct}
{\sc \au{Trias, F.~X.}, \au{Soria, M.}, \au{Oliva, A.} \&
  \au{P{\'e}rez-Segarra, C.~D.}} \yr{2007}  \at{{Direct numerical simulations
  of two-and three-dimensional turbulent natural convection flows in a
  differentially heated cavity of aspect ratio 4}}.  \jt{J. Fluid Mech.}
  \bvol{586},  \pg{259--293}.

\bibitem[de~Vahl~Davis \& Jones(1983)]{de1983natural}
{\sc \au{de~Vahl~Davis, G.} \& \au{Jones, I.~P.}} \yr{1983}  \at{{Natural
  convection in a square cavity: a comparison exercise}}.  \jt{Int. J. Numer.
  Methods Fluids}  \bvol{3},  \pg{227--248}.

\bibitem[Verzicco \& Orlandi(1996)]{verzicco1996finite}
{\sc \au{Verzicco, R} \& \au{Orlandi, P}} \yr{1996}  \at{{A finite-difference
  scheme for three-dimensional incompressible flows in cylindrical
  coordinates}}.  \jt{J. Comput. Phys.}  \bvol{123},  \pg{402--414}.

\bibitem[Wang {\em et~al.\/}(2020{\natexlab{{\em a\/}}})Wang, Chong, Stevens,
  Verzicco \& Lohse]{wang2020zonal}
{\sc \au{Wang, Q.}, \au{Chong, K.-L.}, \au{Stevens, R. J. A.~M.}, \au{Verzicco,
  R.} \& \au{Lohse, D.}} \yr{2020{\natexlab{{\em a\/}}}}  \at{{From zonal flow
  to convection rolls in Rayleigh-B\'enard convection with free-slip plates}}.
  \jt{J. Fluid. Mech.}  \bvol{905},  \pg{A21}.

\bibitem[Wang {\em et~al.\/}(2020{\natexlab{{\em b\/}}})Wang, Shishkina \&
  Lohse]{wang2020ihc}
{\sc \au{Wang, Q.}, \au{Shishkina, O.} \& \au{Lohse, D.}}
  \yr{2020{\natexlab{{\em b\/}}}}  \at{Scaling in internally heated convection:
  a unifying theory}.  \jt{Geophys. Res. Lett.(submitted),arXiv:2010.05789} .

\bibitem[Wang {\em et~al.\/}(2020{\natexlab{{\em c\/}}})Wang, Verzicco, Lohse
  \& Shishkina]{wang2020multiple}
{\sc \au{Wang, Q.}, \au{Verzicco, R.}, \au{Lohse, D.} \& \au{Shishkina, O.}}
  \yr{2020{\natexlab{{\em c\/}}}}  \at{Multiple states in turbulent
  large-aspect ratio thermal convection: What determines the number of
  convection rolls?}  \jt{Phys. Rev. Lett.}  \bvol{125},  \pg{074501}.

\bibitem[Wang {\em et~al.\/}(2018{\natexlab{{\em a\/}}})Wang, Wan, Yan \&
  Sun]{wang2018multiple}
{\sc \au{Wang, Q.}, \au{Wan, Z.-H.}, \au{Yan, R.} \& \au{Sun, D.-J.}}
  \yr{2018{\natexlab{{\em a\/}}}}  \at{{Multiple states and heat transfer in
  two-dimensional tilted convection with large aspect ratios}}.  \jt{Phys. Rev.
  Fluid}  \bvol{3},  \pg{113503}.

\bibitem[Wang {\em et~al.\/}(2018{\natexlab{{\em b\/}}})Wang, Xia, Wang, Sun,
  Zhou \& Wan]{wang2018flow}
{\sc \au{Wang, Q.}, \au{Xia, S.-N.}, \au{Wang, B.-F.}, \au{Sun, D.-J.},
  \au{Zhou, Q.} \& \au{Wan, Z.-H.}} \yr{2018{\natexlab{{\em b\/}}}}  \at{Flow
  reversals in two-dimensional thermal convection in tilted cells}.  \jt{J.
  Fluid Mech.}  \bvol{849},  \pg{355--372}.

\bibitem[Wang {\em et~al.\/}(2019)Wang, Xia, Yan, Sun \& Wan]{wang2019non}
{\sc \au{Wang, Q.}, \au{Xia, S.-N.}, \au{Yan, R.}, \au{Sun, D.-J.} \& \au{Wan,
  Z.-H.}} \yr{2019}  \at{{Non-Oberbeck-Boussinesq effects due to large
  temperature differences in a differentially heated square cavity filled with
  air}}.  \jt{Int. J. Heat Mass Transf.}  \bvol{128},  \pg{479--491}.

\bibitem[Xia(2013)]{xia2013current}
{\sc \au{Xia, K.-Q.}} \yr{2013}  \at{Current trends and future directions in
  turbulent thermal convection}.  \jt{Theor. Appl. Mech. Lett.}  \bvol{3},
  \pg{052001}.

\bibitem[Xin \& Le~Qu{\'e}r{\'e}(1995)]{xin1995direct}
{\sc \au{Xin, S.} \& \au{Le~Qu{\'e}r{\'e}, P.}} \yr{1995}  \at{Direct numerical
  simulations of two-dimensional chaotic natural convection in a differentially
  heated cavity of aspect ratio 4}.  \jt{J. Fluid Mech.}  \bvol{304},
  \pg{87--118}.

\bibitem[Yano {\em et~al.\/}(2003)Yano, Talagrand \& Drossart]{yano2003outer}
{\sc \au{Yano, J.-I.}, \au{Talagrand, O.} \& \au{Drossart, P.}} \yr{2003}
  \at{{Outer planets: Origins of atmospheric zonal winds}}.  \jt{Nature}
  \bvol{421},  \pg{36}.

\bibitem[Zhang {\em et~al.\/}(2020)Zhang, Van~Gils, Horn, Wedi, Zwirner,
  Ahlers, Ecke, Weiss, Bodenschatz \& Shishkina]{zhang2020boundary}
{\sc \au{Zhang, X.}, \au{Van~Gils, D. P.~M.}, \au{Horn, S.}, \au{Wedi, M.},
  \au{Zwirner, L.}, \au{Ahlers, G.}, \au{Ecke, R.~E.}, \au{Weiss, S.},
  \au{Bodenschatz, E.} \& \au{Shishkina, O.}} \yr{2020}  \at{{Boundary zonal
  flow in rotating turbulent Rayleigh-B{\'e}nard convection}}.  \jt{Phys. Rev.
  Lett.}  \bvol{124},  \pg{084505}.

\bibitem[Zwirner {\em et~al.\/}(2020)Zwirner, Khalilov, Kolesnichenko, Mamykin,
  Mandrykin, Pavlinov, Shestakov, Teimurazov, Frick \&
  Shishkina]{zwirner2020influence}
{\sc \au{Zwirner, L.}, \au{Khalilov, R.}, \au{Kolesnichenko, I.}, \au{Mamykin,
  A.}, \au{Mandrykin, S.}, \au{Pavlinov, A.}, \au{Shestakov, A.},
  \au{Teimurazov, A.}, \au{Frick, P.} \& \au{Shishkina, O.}} \yr{2020}  \at{The
  influence of the cell inclination on the heat transport and large-scale
  circulation in liquid metal convection}.  \jt{J. Fluid Mech.}  \bvol{884},
  \pg{A18}.

\bibitem[Zwirner \& Shishkina(2018)]{zwirner2018confined}
{\sc \au{Zwirner, L.} \& \au{Shishkina, O.}} \yr{2018}  \at{{Confined inclined
  thermal convection in low-Prandtl-number fluids}}.  \jt{J. Fluid Mech.}
  \bvol{850},  \pg{984--1008}.

\end{thebibliography}

\end{document}